\documentclass[11pt]{article}
\usepackage{fullpage}
\usepackage{verbatim}
\usepackage{amsmath,amssymb,amsthm}
\usepackage[style=alphabetic,natbib=true,maxbibnames=99,minalphanames=3,maxalphanames=4]{biblatex}
\usepackage{hyperref}
\usepackage{xcolor}

\usepackage[linesnumbered,ruled,vlined]{algorithm2e}

\newtheorem{theorem}{Theorem}
\newtheorem{lemma}{Lemma}
\newtheorem{corollary}{Corollary}
\newtheorem{proposition}{Proposition}
\theoremstyle{definition}

\newcommand{\cF}{\mathcal{F}}

\newcommand{\bE}{\mathbb{E}}
\newcommand{\OPT}{\mathsf{OPT}}
\newcommand{\opt}{\mathsf{opt}}

\newcommand{\ALG}{\mathsf{ALG}}
\newcommand{\alg}{\mathsf{alg}}
\newcommand{\argmax}{\operatorname*{argmax}}

\begin{document}

\title{Optimal Prophet Inequality for Unions of Easy Systems}
\author{
  Hanrui Zhang
  \qquad
  Yiran Zhang
  \\[1ex]
  \normalsize Chinese University of Hong Kong \\
  \normalsize \texttt{hanrui@cse.cuhk.edu.hk, yiranzhang@link.cuhk.edu.hk}
}
\date{}

\maketitle
\thispagestyle{empty}

\begin{abstract}
We give an $O(\log k / \log \log k)$ prophet inequality when the system of feasible sets is the union of $k$ ``easy'' systems (e.g., matroids and knapsacks) each admitting a constant-factor prophet inequality.
This implies an $O(\log k / \log \log k)$ prophet inequality for downward-closed systems with $k$ maximal feasible sets, matching the lower bound by \citet{babaioff2007matroids}.
As a result, under the parametrization by $k$, our upper bound is asymptotically optimal for downward-closed systems, and by implication, unions of $k$ easy systems.
\end{abstract}

\clearpage
\setcounter{page}{1}

\section{Introduction}

Prophet inequalities are algorithms that, given distributional information, select items online subject to feasibility, so as to maximize the total value of items selected.
More specifically, a total of $n$ items have independent values, whose distributions are known to the algorithm upfront.
Items arrive online one by one in an adversarial order.
Upon the arrival of each item, the algorithm observes the realized value of the item, and must decide immediately and irrevocably whether to select the item.
Throughout the procedure, the items selected must always be feasible, i.e., the set of items selected must be a member of a prescribed system of feasible sets.
Different choices of the system of feasible sets lead to natural special cases of the problem, including single-choice prophet (the system consists of all singletons), multi-unit prophet (the system consists of all sets below a certain size), matroid prophet (the system corresponds to independent sets of a given matroid), etc.

As a result of persistent interest and effort in prophet inequalities, most important variants have now been largely resolved, in the sense that we know at least the asymptotically tight competitive ratio (i.e., value achievable online as a fraction of the offline optimum), and often even the precise numerical value thereof.
Super-constant gaps remain in very few cases, among which perhaps the cleanest one is prophet inequalities with {\em general downward-closed systems} of feasible sets: \citet{rubinstein2016beyond} presents an $O(\log^2 n)$ upper bound,\footnote{To be more precise, the bound is $O(\log n \cdot \log r)$, where $r$ is the size of the largest feasible set.} while \citet{rubinstein2026secretary} give a lower bound of $\Omega((\log n / \log \log n)^2)$.
The two bounds almost match each other, but not exactly.\footnote{Very recently, \citet{zhang2026tight} established matching upper and lower bounds of $\Theta(\log^2n/\log \log n)$.}
This gap, while small, motivates our work in a curious sense discussed below --- in fact, as a corollary of our main result, we will present a tight bound for the same problem under a different parametrization.  

The above bounds are parametrized by $n$, the number of items.
We take another perspective, which, in the case of downward-closed systems, involves parametrizing the problem by $k$, the number of maximal feasible sets.
While less popular, this is not unreasonable --- for example, \citet{feldman2025bayesian} parametrize their bound using the same parameter in the context of clock auctions.
More importantly, this parametrization enables a new, more general interpretation of downward-closed systems: A downward-closed system with $k$ maximal sets is the union of $k$ power sets (or free matroids), each generated by one maximal set.
Under this interpretation, downward-closed systems generated by $k$ maximal sets become a {\em special case} of something highly general: For example, one could consider systems that are unions of $k$ matroids, or $k$ sub-systems each defined by a knapsack constraint, or even mixtures thereof.
In fact, our main result applies not only to systems with at most $k$ maximal sets, but also more generally to any system that is the union of $k$ {\em easy systems}, each admitting a (for example) constant-factor prophet inequality on its own.
In other words, we measure the complexity of a problem instance by not how many items, but how many disjunctive easy systems / constraints it involves.

\subsection{Results and Techniques}

Our main result is a generic algorithmic reduction, from a problem instance defined on the union of $k$ easy systems, to those defined on each of these easy systems:

\begin{theorem}[informal version of Theorem~\ref{thm:maintheorem}]
    Consider $k$ systems $\cF_1, \dots, \cF_k$ defined over a common ground set of independent items.
    Suppose there is some $\alpha \ge 1$ such that for each $i \in [k]$, the problem instance defined by $\cF_i$ alone admits an $\alpha$-competitive prophet inequality.
    Then the problem instance defined by $\cF = \bigcup_{i \in [k]} \cF_i$ admits an $O(\alpha \cdot \log k / \log \log k)$-competitive prophet inequality, where $O(\cdot)$ hides an absolute constant.
\end{theorem}

As a direct corollary, there exists an $O(\log k / \log \log k)$-competitive prophet inequality for any problem instance defined by the union of $k$ systems each admitting a constant-factor prophet inequality, including matroids, knapsacks, etc.
This in particular subsumes downward-closed systems as a special case, which are unions of free matroids.
As a result, we obtain the following bound for downward-closed systems generated by $k$ maximal sets:

\begin{corollary}[informal version of Corollary~\ref{cor:feasiblesets}]
    Any problem instance defined by a downward-closed system containing $k$ maximal sets admits an $O(\log k / \log \log k)$-competitive prophet inequality.
\end{corollary}

Note that this bound is {\em asymptotically tight}: The hard instance by \citet{babaioff2007matroids} (in which $k = \mathrm{poly}(n)$) rules out $o(\log k / \log \log k)$-competitive prophet inequalities even restricted to the case of disjoint maximal sets and iid items with binary values.
By implication, our main result for unions of $k$ easy systems is also tight under the $k$-parametrization.

\begin{proposition}[informal, corollary of \citep{babaioff2007matroids}]
    The above upper bounds for downward-closed systems and unions of easy systems are asymptotically optimal under the $k$-parametrization.
\end{proposition}

In fact, our bound for downward-closed systems is the first asymptotically tight bound, albeit under a less canonical parametrization.
While this bound does not seem to easily imply an improved bound under the $n$-parametrization, it (as well as the underlying technical insights explained below) does hint at the possibility of such improvements.\footnote{
    Superficially, to obtain a better bound than $O(\log^2 n)$, one would need $k = \exp(o(\log^2 n \log \log n))$, whereas in general $k$ could be as large as $\exp(\Omega(n))$ (e.g., a system containing all sets of size precisely $n / 2$ and all subsets thereof).
    Certain more complex ideas, e.g., trying to cover a system using a few matroids (which in particular handles the above ``hard'' instance), do not easily give better bounds either.
}
At the same time, it narrows down the space of potential stronger lower bound constructions: We know that any improved lower bound construction would necessarily involve a large number of maximal feasible sets.
On top of that, it must be structurally very rich, in the sense that it cannot be written as the union of polynomially many easy systems.

\paragraph{Overview of algorithm and analysis.}
The reduction itself is incredibly clean.
We first rule out two easy cases:
\begin{itemize}
    \item With constant probability, a single item contributes a significant (i.e., $\Omega(\log \log k / \log k)$) fraction of the offline optimum benchmark (call it $\opt$).
    \item Items with large realized values (we will define ``large'' appropriately so that large realized values are rare) contribute more than half of $\opt$ in expectation.
\end{itemize}
In both cases, we simply take the first item that is large enough.
In the remaining nontrivial case, we essentially pick the ``most promising'' sub-system out of the $k$, which is most likely to contain a set whose value is at least a constant fraction of $\opt$, and disregard all the other possibilities.
Restricted to the most promising sub-system, we run the existing $\alpha$-competitive algorithm as a blackbox subroutine.
This works because of concentration\footnote{
    Here, concentration requires $\opt$ to be distributed somewhat evenly among sufficiently many small items.
    This can possibly be violated only in the easy cases above, which are handled separately.
}, as we explain below.

Let $v_1, \dots, v_n$ be the values of the $n$ items, which are independent random variables.
Consider the maximum value, say $X$, generated by the best set in any given system $\cF$ (which could be either the union system or a sub-system, depending on the context), i.e.,
\[
    X = \max_{S \in \cF} \sum_{i \in S} v_i.
\]
First, imagine $\cF$ is the union system.
\citet{rubinstein2016beyond} already observes that $X$ is reasonably likely to be large enough, i.e., with constant probability, $X$ is at least a constant fraction of its expectation ($\opt$ in this case).
Given this, applying a simple union bound over the $k$ sub-systems, we see that with probability $\Omega(1/k)$, the most promising sub-system generates at least a constant fraction of $\opt$.
That is, the most promising system is at least {\em somewhat likely to be very good}.
Below we argue that this implies that the most promising system must be {\em somewhat good in expectation}, i.e., it generates a substantial fraction (i.e., $\Omega(\log \log k / \log k)$) of $\opt$ in expectation.

Now imagine $\cF$ is our most promising system.
We observe (as a corollary of well-established statistical facts) that $X$ (which is now the value generated by the most promising system) is rarely much larger than its expectation.
In fact, as long as $\bE[X]$ is not too small (which, as we argue, must be the case), the tail probability $\Pr[X \ge t \cdot \bE[X]]$ for $t \ge 1$ decays {\em super-exponentially} in $t$:
\[
    \Pr[X \ge t \cdot \bE[X]] \le \exp(-\Omega(t \log t)).
\]
So, the probability that $X$ is more than $\bE[X] \cdot C \log k / \log \log k$ for some sufficiently large constant $C$ must be sufficiently small (i.e., $o(1 / k)$).
On the other hand, recall that the probability that $X$ is a constant fraction of $\opt$ is $\Omega(1 / k)$, which means $\opt \le C \log k / \log \log k \cdot \bE[X]$.
Now if we run an $\alpha$-competitive algorithm for the most promising sub-system, the value we get would be $\bE[X] / \alpha = \opt \cdot \log \log k / (\alpha \cdot C \log k)$.
In other words, in addition to the factor of $\alpha$, we lose a factor of $O(\log k / \log \log k)$ in the entire process.

\subsection{Further Related Work}

Prophet inequalities have been extremely well-studied in different models and contexts. 
In the classical single-choice prophet inequality, \citet{krengel1977semiamarts} show that the online algorithm can achieve at least half of the expected offline optimum. 
Subsequent research has explored the prophet problem in a variety of settings, including matroids \citep{kleinberg2012matroid,feldman2016online,lee2018optimal}, knapsack constraints \citep{feldman2015asimple,dutting2017prophet,jiang2022tight}, polymatroids \citep{dutting2015polymatroid}, $k$-choice (uniform matroids) \citep{hajiaghayi2007automated,alaei2011bayesian,jiang2022tight}, matching constraints \citep{gravin2019prophet,ezra2020online}, independent sets in graphs \citep{gobel2014online}, combinatorial allocation \citep{feldman2015combinatorial,dutting2020prophet,dutting2020anprophet,zhang2022improved,correa2023aconstant}, single-sample algorithms \citep{azar2014prophet,rubinstein2020optimal,caramanis2022single,dutting2024online}, etc. 
For comprehensive coverage of the problem and its variants, we refer interested readers to surveys \citep{lucier2017aneconomic,correa2018recent}.

Having set the broader context, we now turn to the lines of research most closely related to the present work. 
For prophet inequalities on general downward-closed set systems, in addition to the aforementioned papers \citep{babaioff2007matroids,rubinstein2016beyond,rubinstein2026secretary}, \citet{rubinstein2017combinatorial} extend the problem to combinatorial valuation functions. 
This research is also relevant to intersections and unions of matroid systems. 
\citet{kleinberg2012matroid} show a competitive ratio of $(4k - 2)$ for the intersection of $k$ matroids and a lower bound of $\Omega(\sqrt{k})$. \citet{feldman2016online} demonstrate an improved upper bound of $e (k + 1)$, \citet{saxena2023animproved} establish an improved lower bound of $k^{1 / 2 + \Omega(1 / \log \log k)}$, which \citet{fotakis2026nearly} subsequently improve to $\frac{k}{4\log(1+(e-1)k)}$.
\citet{alon2025abicriterion} consider ``$k$-fold matroid unions'', and present a prophet inequality which achieves at least $( 1 - O(\sqrt{(\log k) / k}) )$ of the expected optimum.
This is different from unions of $k$ matroids that we consider: The $k$-fold union of matroids is defined by taking one independent set from each of the $k$ matroids, putting the union of these independent sets into the new system, and enumerating all combinations of independent sets.
As a result, the problem defined by the $k$-fold union of matroids becomes {\em easier}.
To the best of our knowledge, no prior work has studied prophet inequalities for unions of easy systems, or even special cases thereof, e.g., unions of matroid independent systems.
Our main result extends the problem to a broader class of constraints.

\section{Preliminaries}

\paragraph{The downward-closed prophet problem.}
Consider a collection of $n$ items $[n]$. 
Each item $i$ has a value $v_i$, drawn independently from a known distribution $D_i$ over $\mathbb{R}_{\ge 0}$. 
Let $\cF$ be an arbitrary downward-closed set system on the item set. 
The algorithm has prior knowledge of $n$, $\cF$, and all distributions $\{ D_i \}$, and maintains a set $\ALG$ of selected elements.\footnote{
    Slightly abusing notation, we will use $\ALG$ to denote both the algorithm and the set of items it selects.
}
An (adaptive online) adversary determines the arrival order: At each time, some unseen item $i$ arrives, and its value $v_i \sim D_i$ realizes.
Upon observing the realized value $v_i \sim D_i$, the algorithm must make an immediate and irreversible decision about whether to include the corresponding item in $\ALG$, while keeping $\ALG$ feasible (i.e., $\ALG \in \cF$). 
The objective is to maximize (the expectation of) the total selected value, i.e.,
\[
    \alg = \bE \left[\sum_{i \in \ALG} v_i\right].
\]

Let $\OPT := \argmax_{S \in \cF} \sum_{i \in S} v_i$ denote the optimal feasible set in hindsight (ties broken arbitrarily), and
\[
    \opt := \bE\left[\sum_{i \in \OPT} v_i\right] = \bE \left[ \max_{S \in \cF} \sum_{i \in S} v_i \right]
\]
denote the expected offline optimum.
We say that an online algorithm induces a prophet inequality with competitive ratio $\beta \ge 1$, if for any distributions $\{ D_i \}_{i = 1}^n$,
\[
    \beta \cdot \alg = \beta \cdot \bE \left[ \sum_{i \in \ALG} v_i \right]
    \ge \bE \left[ \max_{S \in \cF} \sum_{i \in S} v_i \right] = \opt,
\]
where the expectations are taken over the randomness in the realization process (and possibly the randomness of the algorithm).

\paragraph{The system of feasible sets as the union of $k$ easy systems.}

We say that a set system $\cF$ is the union of $k$ independence systems, iff there exist independence systems $\cF_1, \cdots, \cF_k \subseteq 2^{[n]}$ such that $\cF = \cup_{j \in [k]} ~ \cF_j$. 
In other words, a set $S$ is feasible in $\cF$ iff there exists some $j \in [k]$ such that $S \in \cF_j$. 

As briefly discussed earlier, considering the union of $k$ easy systems is more general than that of $k$ maximal feasible sets. 
Since each system $\cF_i \subseteq 2^{[n]}$ is a collection of maximal sets, it may, for example, be a matroid or a knapsack constraint. 
Also note that in principle, the $k$ sub-systems are not necessarily part of the problem instance, but something the algorithm is free to choose.
For example, one could in principle view the collection of independent sets of a matroid as the union of the power sets generated by all its maximal members, where each maximal member induces its own easy sub-system.
However, treating the entire collection of independent sets as one easy system, in our framework, normally leads to a better overall competitive ratio.

\section{Algorithm for Unions of Easy Systems}

In this section, we assume that for each $j \in [k]$, there exists a prophet algorithm $\ALG_j$ for the system $\cF_j$ that gives a competitive ratio of $\alpha$.
For ease of presentation, we assume items arrive from $1$ to $n$ (our algorithm does not utilize the order of arrival).
We begin with a verbal description of the algorithm:

\begin{enumerate}
\item Initialize $\ALG \gets \emptyset$. 
Set $t$ as the largest number such that $\Pr [\max_i v_i \ge t] \ge \frac{1}{2}$.

\item If $t > (\opt \cdot \log \log k)/(200000 \log k)$, take the first item with a value greater than or equal to $t$. 

\item 
Otherwise, let $\opt_{> 2t}$ denote the expected contribution from values greater than $2t$, i.e.,
\[
    \opt_{> 2t} := \bE\left[\sum_{i \in \OPT: v_i > 2t} v_i\right].
\]
If $\opt_{> 2t} \ge \frac{\opt}{2}$, then take the first item with a value $> 2t$.

\item
Otherwise, ignore values greater than $2t$, and let $\opt_{\le 2t}$ denote the expected offline optimum of the resulting instance. 
That is,
\[
    \opt_{\le 2t} := \bE\left[\max_{S \in \cF} \sum_{i \in S} v_i^{\le 2t}\right],
\quad \text{where} \quad
    v_i^{\le 2t} = \begin{cases}
        v_i, & \text{if } v_i \le 2t \\
        0, & \text{otherwise}
    \end{cases}.
\]
Choose the independence system $\cF_j$ that maximizes $\Pr [\max_{S \in \cF_j} \sum_{i \in S} v_i \ge \frac{\opt_{\le 2t}}{2}]$.
Choose the set $\ALG$ by applying the existing $\alpha$-competitive algorithm $\ALG_j$ to $\cF_j$.
(This guarantees an expected value no less than $\frac{1}{\alpha} \cdot \bE [\max_{S \in \cF_j} \sum_{i \in S} v_i]$.)
\end{enumerate}

The formal description is given as Algorithm~\ref{alg}.
We argue that this algorithm admits a competitive ratio of $O(\alpha \cdot \log k / \log \log k)$.

\begin{theorem}
\label{thm:maintheorem}
    Consider $k$ independence systems $\cF_1, \cdots, \cF_k$ defined over $[n]$. 
    Suppose that for each $j \in [k]$, the problem instance defined by $\cF_j$ alone admits an $\alpha$-competitive prophet inequality. 
    Then Algorithm \ref{alg} is $O(\alpha \cdot \log k / \log \log k)$-competitive for the instance defined by $\cF = \cup_{j \in [k]} \cF_j$.
\end{theorem}

Theorem \ref{thm:maintheorem} is proved in Sections \ref{sec:easy}--\ref{sec:summary}. 
Before proceeding to the proof, we state the following corollary of Theorem~\ref{thm:maintheorem}.

\begin{corollary}
    \label{cor:feasiblesets}
    For any downward-closed prophet instance containing $k$ maximal feasible sets, Algorithm \ref{alg} obtains a competitive ratio of $O(\log k / \log \log k)$.
\end{corollary}
\begin{proof}[Proof of Corollary \ref{cor:feasiblesets} given Theorem \ref{thm:maintheorem}]
    Each maximal feasible set is a free matroid. For each $j \in [k]$, take $\cF_j$ as the power set of the $j$-th maximal feasible set $M_j \in \cF$, we have $\cF = \cup_{j \in [k]} \cF_j$. 
    For each prophet instance defined by $\cF_j$, selecting all items in the set (i.e., $\ALG_j = M_j$ for each $j \in [k]$) gives a $1$-competitive prophet inequality. 
    By Theorem \ref{thm:maintheorem}, Algorithm \ref{alg} yields a competitive ratio of $O(\log k / \log \log k)$.
\end{proof}

\begin{algorithm}[t]
\caption{Algorithm for unions of $k$ easy systems.}
\label{alg}
\KwIn{$n$, downward-closed set system $\cF$, and distributions $\{ D_i \}_{i = 1}^n$}
\KwOut{Set of items $\ALG$}
\BlankLine
\tcc{Assume for each $j \in [k]$, there exists an $\alpha$-competitive algorithm $\ALG_j$ for the system $\cF_j$.}
Initialize $\ALG \gets \emptyset$\;
$t \gets \max\{ x : \Pr [\underset{i}{\max} ~ v_i \ge x] \ge \frac{1}{2} \}$\;
\BlankLine
\uIf{$t > \frac{\opt \cdot \log \log k}{200000 \log k}$}{
    $i^* \gets \min \{ i : v_i \ge t \}$\;
    $R \gets \{ i^* \}$\;
}
\uElseIf{$\opt_{> 2t} \ge \frac{\opt}{2}$}{
    $i^* \gets \min \{ i : v_i > 2t \}$\;
    $R \gets \{ i^* \}$\;
}
\Else{
    Ignore values $> 2t$\;
    $\cF_{j^*} \gets \underset{\cF_j \subseteq \cF}{\arg \max} ~ \Pr \left[ \underset{S \in \cF_j}{\max} \sum_{i \in S} v_i \ge \frac{\opt_{\le 2t}}{2} \right]$\;
    $R \gets \ALG_{j^*} (\cF_{j^*})$\;
}
$\ALG \gets \ALG \cup R$\;
\Return $\ALG$\;
\end{algorithm}

\subsection{The Easy Cases}
\label{sec:easy}

As introduced earlier, there are two easy cases where the algorithm takes the first item that is large enough. 
We summarize the results in the following lemma.

\begin{lemma}
    \label{lem:easycases}
    Consider any downward-closed prophet instance defined as the union of $k$ independence systems.  
    Algorithm \ref{alg} attains a competitive ratio of $O(\log k / \log \log k)$ for the following two cases:
    \begin{enumerate}
        \item $t > \frac{\opt \cdot \log \log k}{200000 \log k }$.
        \item $\opt_{> 2t}$ is no less than $\frac{\opt}{2}$.
    \end{enumerate}
\end{lemma}
\begin{proof}
    Recall that $t$ is the largest number such that $\Pr [\underset{i}{\max} ~ v_i \ge t] \ge \frac{1}{2}$.
    For case \textit{1.}, the probability of taking an item with a value $\ge t$ is at least $\frac{1}{2}$.
    Therefore, the algorithm obtains an expected value of at least $(\opt \cdot \log \log k)/(400000 \log k)$.

    For case \textit{2.}, the algorithm takes the first item with a value $> 2t$.
    The probability of taking an item is less than $\frac{1}{2}$. 
    In other words, for any item $i$, the probability that the algorithm reaches $i$ without selecting any item is greater than $\frac{1}{2}$. 
    We have:
    \begin{align*}
        \mathsf{alg} \ge & \sum_{i = 1}^n \Pr [\text{reaching item } i] \cdot \Pr [v_i > 2t] \cdot \bE [v_i |v_i > 2t] \\
        > & \sum_{i = 1}^n \frac{1}{2} \cdot \bE [v_i \cdot \mathbf{1}_{v_i > 2t}] \\
        = & \frac{1}{2} \sum_{i = 1}^n \bE [v_i \cdot \mathbf{1}_{v_i > 2t}].
    \end{align*}
    $\opt_{> 2t}$ is no greater than $\sum_{i = 1}^n \bE [v_i \cdot \mathbf{1}_{v_i > 2t}]$. 
    Therefore, the algorithm achieves at least a constant fraction of the expected contribution from values $> 2t$, and thus a constant fraction of $\opt$.
\end{proof}

\subsection{Tail Behavior of the Maximum Value}
\label{sec:tailbehavior}
 
Our proof relies on the following tail bound by \cite{Ledoux1997On}:

\begin{theorem}[Theorem 2.4 in \citep{Ledoux1997On}]
    \label{thm:Ledoux}
    Let $Y_i$'s be independent random variables with values in some space $S$; let $\mathcal{C}$ be a countable class of measurable functions $f:S \to [0,1]$; and let $Z = \sup_{f \in \mathcal{C}} \sum_{i = 1}^n f(Y_i)$. 
    Then, for every $t \ge 0$, 
    \begin{equation}
        \label{eq:Ledouxbound}
        \Pr \left[ Z \ge \bE [Z] + t \right] \le \exp \left( -\frac{t}{K} \log \frac{\bE [Z] + t}{\bE [Z]} \right).
    \end{equation}
    For some numerical constant $K>0$.
\end{theorem}

We interpret this bound similarly to \cite{rubinstein2016beyond}. 
Let $Y_i$ be the vector in $[0, 1]^{|\cF_j|}$ whose $S$-th coordinate is $v_i$ if $i \in S$, and $0$ otherwise. 
Let $f_S (Y_i) := [Y_i]_S$ (i.e., the $S$-th coordinate of $Y_i$), so $\sum_{i = 1}^n f_S (Y_i)$ is the value of set $S$. Let $\mathcal{C} := \{ f_S \}_{S \in \cF_j}$. 
And $Z = \sup_{f_S \in \mathcal{C}} \sum_{i = 1}^n f_S (Y_i) = \max_{S \in \cF_j} \sum_{i \in S} v_i$.
We derive the following lemma.

\begin{lemma}
    \label{lem:concentration}
    Consider an independence system $\cF_j$ and independent variables $v_i \in [0,1]$. Let $Z = \max_{S \in \cF_j} \sum_{i \in S} v_i$ with $\bE [Z] = \lambda$. For $x \ge 2$, it holds that:
    \[
        \Pr \left[ Z \ge x \cdot \lambda \right] 
        \le \exp \left( - \frac{1}{6590} \lambda \cdot x \log x \right).
    \]
\end{lemma}

We defer the proof of Lemma~\ref{lem:concentration} to Appendix~\ref{app:proofs}.

\subsection{The Main Reduction}

We next present a concentration bound that is almost identical to Lemma 1 in \citep{rubinstein2016beyond} --- essentially a generalization of the bound by \citeauthor{rubinstein2016beyond} to random variables in $[0,1]$. 
Let $V := \max_{S \in \cF} \sum_{i \in S} v_i$ denote the value of the optimal offline solution (which is a random variable).

\begin{lemma}
    \label{lem:1/4-lowerbound}
    For $\opt \ge \Omega(\frac{\log k}{\log \log k})$,
    \[
        \Pr \left[ V \ge \frac{\opt}{2} \right] \ge \frac{1}{4}.
    \]
\end{lemma}
\begin{proof}
    We have,
    \begin{align*}
        \opt & = \int_0^\infty \Pr \left[ V \ge u \right] \mathrm{d} u  \\
        & = \int_{- \opt}^\infty \Pr \left[ V \ge \opt + t \right] \mathrm{d} t \\
        & = \int_{- \opt}^{- \opt / 2} \Pr \left[ V \ge \opt + t \right] \mathrm{d} t 
        + \int_{- \opt / 2}^\opt \Pr \left[ V \ge \opt + t \right] \mathrm{d} t
        + \int_\opt^\infty \Pr \left[ V \ge \opt + t \right] \mathrm{d} t.
    \end{align*}
    Setting $t = u - \opt$ gives the second equation. For the first two integrals,
    \begin{align*}
        \int_{- \opt}^{- \opt / 2} \Pr \left[ V \ge \opt + t \right] \mathrm{d} t 
        & \le \int_{- \opt}^{- \opt / 2} 1 \mathrm{d} t \le \frac{\opt}{2}, \\
        \int_{- \opt / 2}^\opt \Pr \left[ V \ge \opt + t \right] \mathrm{d} t 
        & \le \int_{- \opt / 2}^\opt \Pr \left[ V \ge \frac{\opt}{2}  \right] \mathrm{d} t
        \le \frac{3 \opt}{2} \Pr \left[ V \ge \frac{\opt}{2} \right].
    \end{align*}

    For the third part, we apply the concentration bound by \cite{Ledoux1997On} in Theorem \ref{thm:Ledoux}. 
    The interpretation is slightly adjusted.
    Let $Y_i$ be the vector in $[0,1]^{|\cF|}$ whose $S$-th coordinate is $v_i$ if $i \in S$, and $0$ otherwise. 
    Let $f_S(Y_i) := [Y_i]_S$ (i.e., the $S$-th coordinate of $Y_i$), so $\sum_{i = 1}^n f_S (Y_i)$ is the value of set $S$. 
    Let $\mathcal{C} := \{ f_S \}_{S \in \cF }$. 
    And $Z = \sup_{f_S \in \mathcal{C}} \sum_{i = 1}^n f_S (Y_i) = \max_{S \in \cF} \sum_{i \in S} v_i = V$.
    The concentration inequality can be rewritten as:
    \[
        \Pr \left[ V \ge \opt + t \right] 
        \le \exp \left( - \frac{t}{K} \log \left( 1 + \frac{t}{\opt} \right) \right).
    \]

    Therefore, we have,
    \begin{align*}
        \int_{\opt}^{\infty} \Pr \left[ V \ge \opt + t \right] \mathrm{d} t 
        & \le \int_{\opt}^{\infty} \exp \left( - \frac{t}{K} \log \left( 1 + \frac{t}{\opt} \right) \right) \mathrm{d} t \\
        & \le \int_{\opt}^{\infty} \exp \left( - \frac{t}{K} \right) \mathrm{d} t 
        = K \cdot \exp \left( - \frac{\opt}{K} \right).
    \end{align*}

    $K \cdot \exp \left( - \frac{\opt}{K} \right)$ is negligible since $\opt \ge \Omega(\frac{\log k}{\log \log k})$. Collecting the three parts, we get,
    \[
        \opt \le \frac{\opt}{2} + \frac{3\opt}{2} \Pr \left[ V \ge \frac{\opt}{2} \right] + o(1).
    \]

    It follows that,
    \[
        \Pr \left[ V \ge \frac{\opt}{2} \right] \ge \frac{1}{3} - o(1). \qedhere
    \]
\end{proof}

The next lemma establishes that for sufficiently large $\opt$, we can get a constant lower bound on the expected value of the selected sub-system.

\begin{lemma}
    \label{lem:expectation>=1}
    When $v_i \in [0, 1]$ for any item $i$ and $\opt \ge \frac{50000 \log k}{\log \log k}$, for the system $\cF_j$ that maximizes $\Pr[ \max_{S \in \cF_j} \sum_{i \in S} v_i  \ge \frac{\opt}{2} ]$, it holds that $\bE [\max_{S \in \cF_j} \sum_{i \in S} v_i] \ge 1$.
\end{lemma}
\begin{proof}
    Assume, for the sake of contradiction, that $\lambda_j = \bE [\max_{S \in \cF_j} \sum_{i \in S} v_i] < 1$. Apply Lemma \ref{lem:concentration}, we have:
    \begin{align*}
        \Pr \left[ \underset{S \in \cF_j}{\max} \sum_{i \in S} v_i \ge \frac{\opt}{2} \right] 
        & \le \Pr \left[ \underset{S \in \cF_j}{\max} \sum_{i\in S} v_i \ge \frac{25000\log k}{\log \log k} \right] \\
        & \le \exp \left( - \frac{1}{6590} \cdot \frac{25000 \log k}{\log \log k} \log \left( \frac{25000 \log k}{\log \log k} \cdot \frac{1}{\lambda_j} \right) \right) \\
        & < \exp \left( - \frac{1}{6590} \cdot \frac{25000 \log k}{\log \log k} \log \left( \frac{25000 \log k}{\log \log k} \right) \right) \\
        & = \exp \left( - \frac{2500 \log k}{659\log \log k} \left( \log 25000 + \log \log k - \log \log \log k \right) \right) \\
        & = k^{\left( - \frac{2500}{659} \left( \frac{\log 25000}{\log \log k} + 1 - \frac{\log \log \log k}{\log \log k} \right) \right)} \\
        & < k^{\frac{2500}{659} \left( \frac{\log \log \log k}{\log \log k} - 1 \right)} \\
        & \le k^{\frac{2500}{659} \left( \frac{1}{e} - 1 \right)} \\
        & < k^{-2.3}.
    \end{align*}
    The second last inequality is due to $\frac{\log \log \log k}{\log \log k} \le \frac{1}{e}$.

    By Lemma \ref{lem:1/4-lowerbound}, $\Pr[V \ge \frac{\opt}{2}] \ge \frac{1}{4}$. According to the union bound, for the system $\cF_j$ that maximizes $\Pr[ \max_{S \in \cF_j} \sum_{i \in S} v_i  \ge \frac{\opt}{2} ]$, we have,
    \[
        \Pr[\underset{S \in \cF_j}{\max} \sum_{i \in S} v_i  \ge \frac{\opt}{2}] \ge \frac{1}{4k}.
    \]

    This contradicts $\Pr \left[ \max_{S \in \cF_j} \sum_{i \in S} v_i \ge \frac{\opt}{2} \right] < k^{-2.3}$. 
    Therefore, the initial assumption is false, and we conclude that $\bE [\max_{S \in \cF_j} \sum_{i \in S} v_i] \ge 1$.
\end{proof}

It remains to consider the case that $t \le (\opt \cdot \log \log k) / (200000 \log k)$ and $\opt_{> 2t} < \frac{\opt}{2}$. 

\begin{lemma}
    \label{lem:mainreduction}
    Suppose that for each $j \in [k]$, the problem instance defined by $\cF_j$ alone admits an $\alpha$-competitive prophet inequality.
    When $t \le (\opt \cdot \log \log k) / (200000 \log k)$ and $\opt_{> 2t} < \frac{\opt}{2}$, Algorithm \ref{alg} achieves a competitive ratio of $O(\alpha \cdot \frac{\log k}{\log \log k})$.
\end{lemma}
\begin{proof}
    We first show that $\opt_{\le 2t} > \frac{\opt}{2}$. By definition,
    \[
        \opt_{\le 2t} = \bE \left[ \underset{S \in \cF}{\max} \sum_{i \in S} v_i^{\le 2t} \right] \ge \bE \left[ \sum_{i \in \OPT} v_i^{\le 2t} \right] = \opt - \bE \left[ \sum_{i \in \OPT : v_i > 2t} v_i \right] > \frac{\opt}{2}.
    \]

    Recall that we ignore all values greater than $2t$. 
    All items have values in $[0, 2t]$ (i.e., $[0, (\opt \cdot \log \log k) / (100000 \log k)]$).
    Scale the valuation support such that all items have values in $[0,1]$. 
    Let $\{ v_i' \}_{i \in [n]}$ denote the values after scaling; and let $\opt'$ denote the expected offline optimum after scaling:
    \begin{align*}
        v_i' = 
        \begin{cases}
            v_i / \frac{\opt \cdot \log \log k}{100000 \log k}, & \text{if } v_i \le 2t \\
            0, & \text{otherwise}
        \end{cases}
        \qquad \text{and} \qquad
        \opt' = \bE \left[ \underset{S \in \cF}{\max} \sum_{i \in S} v_i' \right].
    \end{align*}

    We get that:
    \[
        \opt' = \frac{\opt_{\le 2t} \cdot 100000 \log k}{\opt \cdot \log \log k} > \frac{50000\log k}{\log \log k}.
    \]

    To show that the algorithm obtains an $\Omega(\frac{1}{\alpha} \cdot \frac{\log \log k}{\log k})$-fraction of $\opt$, it is equivalent to show that after scaling, it gets an $\Omega(\frac{1}{\alpha} \cdot \frac{\log \log k}{\log k})$-fraction of $\opt'$.
    
    Let $\cF_{j^*}$ be the independence system selected by the algorithm that maximizes $\Pr[ \underset{S \in \cF_j}{\max} \sum_{i \in S} v_i'  \ge \frac{\opt'}{2} ]$.
    Let $\gamma$ denote the expected optimum value in $\cF_{j^*}$, i.e., $\gamma := \bE [\max_{S \in \cF_{j^*}} \sum_{i \in S} v_i']$. 
    By Lemma \ref{lem:expectation>=1}, we get $\gamma \ge 1$. 
    
    If $\frac{\opt'}{2 \gamma} < 2$, we would directly get a constant upper bound on $\frac{\opt'}{\gamma}$, which implies a competitive ratio of $O( \alpha )$ for the algorithm. 
    For any $\frac{\opt'}{2 \gamma} \ge 2$, Lemma \ref{lem:concentration} implies that, 
    \begin{align*}
        \Pr \left[ \underset{S \in \cF_{j^*}}{\max} \sum_{i \in S} v_i' \ge \frac{\opt'}{2 \gamma} \cdot \gamma \right] 
        & \le \exp \left( - \frac{\gamma}{6590} \cdot \frac{\opt'}{2 \gamma} \log \left( \frac{\opt'}{2 \gamma} \right) \right)\\
        & \le \exp \left( - \frac{1}{6590} \cdot \frac{\opt'}{2 \gamma} \log \left( \frac{\opt'}{2 \gamma} \right) \right).
    \end{align*}
        
    Let $V'$ denote $\left( \max_{S \in \cF} \sum_{i \in S} v_i' \right)$. Applying Lemma \ref{lem:1/4-lowerbound} yields $\Pr[V' \ge \frac{\opt'}{2}] \ge \frac{1}{4}$. By the union bound, for system $\cF_{j^*}$, we have $\Pr [\max_{S \in \cF_{j^*}} \sum_{i \in S} v_i' \ge \frac{\opt'}{2}] \ge \frac{1}{4k}$. Thus,
    \[
        \frac{1}{4k} \le \Pr \left[ \underset{S \in \cF_{j^*}}{\max} \sum_{i\in S} v_i' \ge \frac{\opt'}{2} \right] 
        \le \exp \left( - \frac{1}{6590} \frac{\opt'}{2 \gamma} \log \frac{\opt'}{2 \gamma} \right).
    \]

    This implies,
    \[
        \frac{\opt'}{2\gamma} = O\left( \frac{\log k}{\log \log k} \right).
    \]

    Applying Algorithm $\ALG_{j^*}$ to system $\cF_{j^*}$ achieves at least $\frac{\gamma}{\alpha}$. Therefore, Algorithm \ref{alg} is $O(\alpha \cdot \frac{\log k}{\log \log k})$-competitive.
\end{proof}

Note that we use the $\alpha$-competitive algorithm $\ALG_{j^*}$ on the selected sub-system $\cF_{j^*}$ as a subroutine. 
Given the existence of constant-factor prophet inequalities on e.g., matroids and knapsacks (\cite{kleinberg2012matroid,dutting2017prophet}), it follows directly that our algorithm provides $O(\frac{\log k}{\log \log k})$-competitive prophet inequalities for unions of matroids, knapsack constraints, or mixtures. 

\subsection{Putting Everything Together}
\label{sec:summary}

Having established the necessary results, we now prove the main theorem.

\begin{proof}[Proof of Theorem \ref{thm:maintheorem}]
    By Lemma \ref{lem:easycases}, we get that the algorithm is $O(\log k / \log \log k)$-competitive for the cases corresponding to line $3$ and line $6$ of Algorithm \ref{alg}.

    Lemma \ref{lem:mainreduction} shows that the algorithm is $O(\alpha \cdot \log k / \log \log k)$-competitive for the case corresponding to line $9$ of Algorithm \ref{alg}. The proof of the theorem is complete.
\end{proof}

\section{A Matching Lower Bound}

To be self-contained, we restate the lower bound construction by \citet{babaioff2007matroids} and show that it gives a lower bound asymptotically matching our upper bound, under the parameterization by $k$. We include the construction stated in Appendix B of \cite{rubinstein2016beyond}.

The idea is the following:
Partition $n$ items into disjoint sets of size $\log n / \log \log n$, each set corresponds to a maximal feasible set in $\cF$. 
Assign values $\{0, 1\}$ independently to the items such that for each item $i$, $v_i = 1$ with probability $\frac{\log \log n}{\log n}$ and $v_i = 0$ with probability $1 - \frac{\log \log n}{\log n}$. 

\begin{theorem}[adapted from \citep{babaioff2007matroids,rubinstein2016beyond}]
    No online algorithm can achieve a competitive ratio of $o(\frac{\log k}{\log \log k})$ for downward-closed prophet.
\end{theorem}
\begin{proof}
    The instance above contains $k = \frac{n \cdot \log \log n}{\log n}$ independence systems. 
    The Theorem in Appendix B of \cite{rubinstein2016beyond} establishes the following conclusion that is relevant to our analysis: $\opt$ is approximately $(\log n / \log \log n)$, and any online algorithm cannot achieve an expected value of more than $2$.
    It is impossible for any online algorithm to achieve a ratio better than $o(\frac{\log n}{\log \log n})$.
    
    We next show that for $\log n > 4$,
    \[
        \frac{\log n}{\log \log n} \ge \frac{\log k}{2 \log \log k}.
    \]
    
    Let $m$ denote $\log n$. We have,
    \[
        \log k = m + \log \log m - \log m \le m.
    \]
    
    For $m > 4$, $m - \log m > m / 2$, we have $\log k > m - \log m > \frac{m}{2}$. Therefore, 
    \[
        \log \log k > \log m - \log 2.
    \]
    
    Thus, $\frac{\log k}{\log \log k} \le \frac{m}{\log m - \log 2}$. 
    For any $m > 4$,
    \[
        \frac{\log n / \log \log n}{\log k / \log \log k} 
        \ge \frac{m}{\log m} \cdot \frac{\log m - \log 2}{m}
        = \frac{\log m - \log 2}{\log m} \ge \frac{1}{2}. \qedhere
    \]
\end{proof}

\printbibliography

\appendix
\section{Omitted Proofs}
\label{app:proofs}

\begin{proof}[Proof of Lemma~\ref{lem:concentration}]
    As shown in the proof of Theorem $2.4$ in \cite{Ledoux1997On}: 
    \begin{align*}
        && \Pr \left[ Z \ge \bE [Z] + t \right] 
        & \le \exp \left( - \frac{t^2}{32 \bE [Z]} \right) & \forall ~ t \le 4 \bE [Z]\phantom{.} \tag{i} \\
        && \Pr \left[ Z \ge \bE [Z] + t \right] 
        & \le \exp \left( - \frac{t}{8} \right) & \forall ~ t > 4 \bE [Z]\phantom{.} \tag{ii} \\
        && \Pr \left[ Z \ge t \right] 
        & \le \exp \left( - \frac{t}{8} \log \frac{t}{\bE [Z]} + t \right) & \forall ~ t \ge 8 \bE [Z]. \tag{iii}
    \end{align*}
    
    To establish an upper bound of the form $\exp \left( - \frac{t}{K} \log \frac{\bE [Z] + t}{\bE [Z]} \right)$ for some constant $K$, we consider the following constraints, depending on the range of $t$
    \footnote{
    Condition (c) is derived from from inequality (iii), whose left-hand side is $\Pr \left[ Z \ge t \right]$; replacing $t$ by $(\bE [Z] + t)$ yields the desired form.
    }:
    \begin{align*}
        && \exp \left( - \frac{t^2}{32 \bE [Z]} \right) 
        & \le \exp \left( - \frac{t}{K} \log \frac{\bE [Z] + t}{\bE [Z]} \right) \qquad & \forall ~ t \le 4 \bE [Z]\phantom{.} \tag{a} \\
        && \exp \left( - \frac{t}{8} \right) 
        &\le \exp \left( - \frac{t}{K} \log \frac{\bE [Z] + t}{\bE [Z]}\right) \qquad & \forall ~ 4 \bE [Z] < t < K \bE [Z]\phantom{.} \tag{b} \\
        && \exp \left( - \frac{t}{8} \log \frac{t}{\bE [Z]} + t \right) 
        &\le \exp \left( - \frac{t - \bE [Z]}{K} \log \frac{t}{\bE [Z]} \right) \qquad & \forall ~ t \ge K \bE [Z]. \tag{c}
    \end{align*}

    We check the three constraints to give a constant lower bound for $K$.
    For condition (a): take the natural logarithm of both sides and divide by $t$. Substituting $\frac{t}{\bE [Z]} = u$, the constraint can be rewritten as:
    \[
        \frac{u}{32} \ge \frac{1}{K} \log(u + 1) \qquad \forall ~ u \in (0,4].
    \]
    That is $K \ge \frac{32 \log(u + 1)}{u}$ for $u \in (0,4]$. Let $f(u) = \frac{32 \log (u + 1)}{u}$. Differentiate,
    \[
        f'(u) = \frac{32 \left( \frac{u}{u + 1} - \log (u + 1) \right)}{u^2}.
    \]
    Let $g(u) = \frac{u}{u + 1} - \log (u + 1)$. Note that $g(0) = 0$.
    \[
        g'(u) = \frac{1}{(u + 1)^2} - \frac{1}{u + 1} <0 \qquad \forall ~ u > 0.
    \]
    Therefore, $f(u)$ is decreasing in $u$. The constraint becomes,
    \[
        K \ge \underset{u \to 0^+}{\lim} f(u) = \underset{u \to 0^+}{\lim} \frac{32 \cdot \frac{1}{u + 1}}{1} = 32.
    \]
    The first equality follows from L'H\^opital's rule.

    For constraint (c), take the natural logarithm of both sides and divide by $t$. Setting $u = \frac{t}{\bE [Z]}$, the constraint can be rewritten as:
    \[
        u \left( \log u - 8 \right) \ge \frac{8 (u - 1)}{K} \log u \qquad \forall ~ u \ge K.
    \]
    Let $h(u) = K u (\log u - 8) - 8(u - 1) \log u$. We need $h(u) \ge 0$ for all $u \ge K$. Take the derivative,
    \begin{align*}
        h'(u) & = K \log u + K - 8K - 8 \log u - 8 + \frac{8}{u} \\
        & > (K - 8) \log u - 7K - 8.
    \end{align*}
    Take $K = e^{8.1}$: for $u \ge K$, $h'(u) > 0$, $h(u)$ is increasing in $u$.
    $h(K) = 0.1 e^{16.2} - 64.8 e^{8.1} +64.8 \approx 8.72 \times 10^5 > 0$. 
    Therefore, taking $K = e^{8.1}$ suffices for the third condition.

    We next consider condition (b) for $t \in (4 \bE [Z], K \bE [Z])$.
    Take the natural logarithm of both sides and divide by $t$.
    Let $u = \frac{t}{\bE [Z]}$, the constraint can be rewritten as:
    \[
        K \ge 8 \log (u + 1) \qquad \forall ~ u \in (4,K).
    \]
    As $8 \log (u + 1)$ is increasing in $u$, $K \ge 8 \log(K + 1)$. For $K = e^{8.1}$, $K \ge 8 \log (e^{8.1} + 1) \approx 64.8$. 

    Combining these cases, taking $K = e^{8.1} \approx 3295$ suffices for the upper bound.

    Rewrite \eqref{eq:Ledouxbound} in terms of $\lambda$ and $x$, we get:
    \[
        \Pr \left[ Z \ge x \cdot \lambda \right] \le \exp \left( - \frac{\lambda}{K} \cdot (x - 1) \log x \right).
    \]
    Take $K = 3295$, for any $x > 1$, 
    \[
        \Pr \left[ Z \ge x \cdot \lambda \right] \le \exp \left( - \frac{\lambda}{3295} \cdot (x - 1) \log x \right).
    \]
    For $x \ge 2$, $\frac{1}{2} \lambda x \log x \le \lambda (x - 1) \log x$. Therefore,
    \[
        \Pr \left[ Z \ge x \cdot \lambda \right] \le \exp \left( - \frac{1}{6590} \lambda \cdot x \log x \right). \qedhere
    \]
\end{proof}

\end{document}